\documentclass[fleqn,usenatbib,useAMS]{mnras}

\usepackage{graphicx}	
\usepackage{amsmath}	
\usepackage{amssymb}	
\usepackage{multicol}        
\usepackage{bm}		
\usepackage{pdflscape}	
\usepackage{hyperref}
\usepackage{esdiff}
\usepackage{natbib} 
\usepackage[outdir=./]{epstopdf}
\usepackage[font=scriptsize]{caption}
\usepackage{xcolor}

\usepackage[T1]{fontenc}
\usepackage{ae,aecompl}

\usepackage{newtxtext,newtxmath}

\title[Slowly rotating accretion flows]{Global structure and dynamics of slowly rotating accretion flows}

\author[R. Ranjbar et al.]{
Razieh Ranjbar,$^{1}$
Amin Mosallanezhad,$^{2}$
Shahram Abbassi $^{1}$ \thanks{E-mail: abbassi@um.ac.ir;}
\\
$^{1}$ Department of Physics, Faculty of Science, Ferdowsi University of Mashhad, Mashhad, 91775-1436, Iran\\
$^{2}$ School of Mathematics and Statistics, Xi'an Jiaotong University, Xi'an, Shaanxi 710049, PR China
}
\date{}

\pubyear{2022}

\begin{document}
\label{firstpage}
\pagerange{\pageref{firstpage}--\pageref{lastpage}}
\maketitle

\begin{abstract}
We study the global solutions of slowly rotating accretion flows around the supermassive black hole in the nucleus of an elliptical galaxy. The velocity of accreted gas surrounding the black hole is initially subsonic and then falls onto the black hole supersonically, so accretion flow must be transonic. We numerically solve equations from the Bondi radius to near the black hole. The focus of our discussion will be on the properties of slightly rotating accretion flows in which radiative losses have been ignored. This study discusses how outer boundary conditions (the temperature and specific angular momentum at the outer boundary) influence accretion flow dynamics. We investigate two physically discontinuous regimes: The Bondi-like type accretion and the Disk-like type accretion. A Bondi-like accretion occurs when the specific angular momentum at the Bondi radius $ \ell_{B} $ is smaller than the specific angular momentum at the marginally stable orbit $ \ell_{ms} $.
In comparison, a Disk-like accretion occurs when the specific angular momentum at the Bondi radius $ \ell_{B} $ is larger than the specific angular momentum of the marginally stable orbit $ \ell_{ms} $. We also keep the assumption of hydrostatic equilibrium and compare our results with the case in which it is not considered. According to this study, considering the assumption of hydrostatic equilibrium reduces the mass accretion rate. Additionally, we find our solution for different ranges of the viscosity parameter $\alpha$. Finally, we study the effect of galaxy potential on slowly rotating accretion flows. 
\end{abstract}




\begingroup
\let\clearpage\relax
\endgroup
\newpage

\section{Introduction}

It has been believed that the supermassive black holes (SMBHs), in the majority 
of nearby galaxies in the universe are powered by hot accretion flows.
For instance, the SMBHs at the center of low-luminosity active galactic nuclei (LLAGNs)
accrete gas in hot accretion mode (see e.g. \cite{2008ARA&A..46..475H}; \cite{2012A&AT...27..557A}). 
Hard/quiescent states of black hole X-ray binaries also operate in hot accretion mode
(see, e.g.\cite{2008NewAR..51..733N}; \cite{2010LNP...794...53B}; \cite{2014ARA&A..52..529Y}).
Compared to the well-known standard thin disks (cold accretion mode), hot accretion disks 
have a much higher temperature and much faster radial velocity. Consequently, 
the radiative efficiency becomes much lower \cite{2014ARA&A..52..529Y}.

There is strong evidence for the coevolution of SMBH at the center of galaxies with 
their host galaxies \cite{2013ARA&A..51..511K}, such as the correlation between the mass 
of the black hole and its luminosity, stellar velocity dispersion, or stellar mass in the 
galaxy bulge (e.g., \cite{2002ApJ...574..740T}; \cite{2009ApJ...698..198G} ). This is known as 
active galactic nucleus (AGN) feedback which plays an essential role in the evolution of 
galaxies. The AGN feedback can change the temperature and density distribution in the
interstellar medium (ISM) in the galaxy with the radiation and outflow (jet/wind), and 
also subsequent changes in black hole fueling and star formation \cite{2012ARA&A..50..455F}; 
\cite{2013ARA&A..51..511K}. Most recent numerical simulations also prove that AGN 
feedback is a common ingredient for our understanding of the evolution of galaxies
\cite{2012ARA&A..50..455F}. 
In terms of the hot accretion flow, AGN feedback can only be in the form of mechanical 
feedback through wind/jet due to the radiative inefficiency of the accretion flows.   

For the study of AGN feedback, an exact evaluation of the mass accretion rate onto the black hole is crucial because it determines the strength of AGN outputs such as wind or jet. Different accretion modes can have different accretion rates. 
To calculate the mass accretion rate in both analytical and numerical simulation works, it is common to adopt the Bondi solution (\cite{1952MNRAS.112..195B}; \cite{2002CUP...1}; \cite{2018MNRAS.479.4778Y}). In the Bondi model, 
the vanishing angular momentum is considered for gas particles, whereas 
in reality, the incoming gas is likely to have a non-negligible rotation motion. Due to the 
existence of angular momentum, the spherical symmetry of the Bondi solution will be broken down. 
Consequently, angular momentum can play a crucial role in the structure and formation of accretion 
flows around black holes.

According to \cite{2001AIPC..556...93Y}, the discrepancy between the mass accretion rate of Sgr A* calculated by the previous ADAF models and one of the hydrodynamical numerical simulations can be reduced if the accretion rotates slowly. So we can conclude that such low angular momentum accretion occurs in various astrophysics systems. For example, observations of elliptical galaxies have revealed a surprising fact, in the densest regions of massive elliptical galaxies the hot gas rotates very slowly (\cite{1977ApJ...218L..43I} and \cite{1980ApJ...241L..65S}). 
Such systems certainly have a giant Bondi-type quasi-spherical accretion flow into their host SMBHs, which extends well beyond the Bondi radius. In recent years, for slowly rotating the gas surrounding the black hole several models have been proposed (\cite{2011MNRAS.415.3721N} hereafter NF11, \cite{2009ApJ...706..637P}). On the other hand, there have been many simulations for slowly rotating accretion flows in the past decades (\cite{2003ApJ...592..767P}, \cite{2019MNRAS.484.1724B} hereafter BY19). \cite{2018MNRAS.476..954Y} revealed that the mass accretion rate drops compared to the Bondi rate when radiation heating and cooling are considered in slowly rotating accretion flows. Also, BY19 studied the slowly rotating low accretion rate flow irradiated by an LLAGN in the presence of wind and the gravity of stars in the galaxy. Further, in recent years, the Bondi model has developed by considering the gravitational potential of the galaxy (\cite{2018ApJ...868...91C}, \cite{2016MNRAS.460.1188K}, \cite{2019MNRAS.489.3870S}, \cite{2000ApJ...528..236Q}) due to the sizeable dynamical scale of the accretion process, from Bondi radius to the Schwarzschild radius of the black hole. 

The Bondi model can be an affordable representative of a hot accretion flow model. 
This is mainly because the observation shows that both the accretion 
flow and AGN outputs themselves are in hot accretion mode. 
In principle, the gravitational energy released in the flow can be carried into the 
the black hole at the center rather than radiated away locally, and some fraction of this 
energy shall be transferred to the jets/wind while the accreting gas moves toward 
the center (see, e.g., \cite{2008NewAR..51..733N};  \cite{2014ARA&A..52..529Y} for 
review of the hot accretion flow model). This model has been applied to the SMBH in 
our Galactic Center, Sagittarius A* (Sgr A*), and M87. 
Since the pioneering work of \cite{1994ApJ...428L..13N} (1995a,b) and \cite{1995ApJ...438L..37A}, 
hot accretion flows have been studied in significant detail both in analytical and numerical 
simulation aspects (\cite{2012ARA&A..50..455F}). In terms of analytical studies, several simplifications 
or assumptions have been considered. In most of those works mentioned above,
the solutions were assumed to follow radial self-similar form, or even for global solutions,
the external boundary conditions of thin accretion disks have been taken into account.

The nature of transonic accretion flows onto a black hole was first introduced by 
\cite{1980ApJ...240..271L} and \cite{1981ApJ...246..314A} where the flow must 
satisfy a regularity condition at the sonic radius. Later on, the structure and stability of transonic 
accretion flows have been investigated by many authors \cite{1989ApJ...336..304A}, \cite{1983AcA....33...79M}. 
For instance, \cite{1997ApJ...476...49N} studied the global structure and Dynamics of the hot 
accretion flow. Their solutions were obtained by solving a set of coupled first-order differential 
equations corresponding to an axisymmetric and steady-state height-integrated accretion flow. 
In this solution, the flow passes through a sonic point and falls supersonically into the SMBH 
with a no-torque condition at the Schwarzschild radius, $ r_{\rm s} $. On the outside, 
at $ 10^{6} r_{\rm s} $, thin accretion disk boundary conditions have been considered. 
In contrast to the above study, \cite{2009ApJ...706..637P}, \cite{2011MNRAS.415.3721N},
focused on a case where the viscous accretion flow located in hot external media
around the Bondi radius. Consequently, the flow will be continuous from the event 
horizon of the SMBH to beyond the Bondi radius. Based on a simple energy consideration,
\cite{2001ApJ...551L..77M} gave a cutting argument against the possible smooth convergence from 
hot accretion flow to standard thin accretion disk (no transition radius).

As mentioned above, NF11 work was different from almost all previous ones in 
the literature of transonic flow solutions in the sense that they considered an axisymmetric, 
steady viscous hot accretion flow extending from beyond the Bondi radius down to the 
black hole event horizon. In contrast to previous studies, they considered very low angular 
momentum for their Bondi-type quasi-spherical accretion flow. Instead of keeping the 
assumption of vertical hydrostatic equilibrium, i.e.,  

\begin{equation} \label{hydro_equilib}
	\frac{1}{\rho} \frac{\partial p}{\partial z} + \frac{\partial \phi}{\partial z} = 0,
\end{equation}
they assumed that both the density $ \rho $ and pressure of the gas $ p $ are 
distributed spherically at each radius down to the black hole (here, $ \phi $ represents  
the gravitational potential). By calculation of the accretion rate at the black hole and  
coupling it to the surrounding gas, enabling AGN feedback to occur; they argued that
the jet power can be a few percent of $ \dot{M}_{\rm B} c^{2} $ where 
$ \dot{M}_{\rm B} $ and $ c $ are Bondi accretion rate and speed of light respectively. 
They did not include the effects of the wind for the estimation of total energy output, 
although in some recent works, the importance of mechanical feedback by wind 
has been recognized (e.g., \cite{2017ApJ...837..149G}; 
\cite{2017MNRAS.465.3291W}, \cite{2017MNRAS.470.4530W}). 

In this study, we revisit Bondi flow from a slowly rotating hot atmosphere by first taking into account a vertical hydrostatic equilibrium equation that is a simple version of the vertical momentum equation (see equation (\ref{H_r}) for more detail). 
This modification can significantly change the accretion rate and consequently 
the total energy output from SMBH. Also, our result indicates that the position of the sonic point, which is one of the eigenvalues in this study, will be changed by this modification. We investigate accretion flow with low angular momentum in two different modes: Bondi-like and Disk-like accretion. Disk-like accretion flows, i.e., those with a small sonic radius, are the focus of the present paper. Compared to NF11, we fixed the outer boundary at the Bondi radius and considered an extensive range of temperatures at the outer boundary. This is because a primary source of the hot interstellar medium in the elliptical galaxy could be mass-loss stellar, which heats the medium to a temperature of around $ 10^6 - 10^7 $ K (\cite{2003A&A...41.1...191}). Therefore, considering a wide range of temperatures at the outer boundary is more reasonable than assuming a single temperature. Moreover, as we know, the gravity of stars might be necessary for regions near the galaxy's center. For that reason, unlike NF11, we include the galactic contribution to the potential in part of the present study and compare our results with and without the inclusion of this potential (see section \ref{Phi} for more detail). Finally, we compare the results of our model with the most updated numerical simulations in the presence of galaxy potential. The effects of outflow as effective feedback are beyond the scope of the paper, and we postpone this issue to our next article.

 The structure of the paper is arranged as follows. We start with analyzing the fundamental equations in \textsection{2}, in \textsection{3} various boundary conditions assumed, and explain our numerical method.
The results of the numerical solution are given in \textsection{4} and finally, \textsection{5} we present related discussions and provide an outline.

\section{Model}
\subsection{Conservation Equations} \label{sec:equations}

We consider transonic, steady, non-self-gravitating, viscous accretion flows with a finite and small angular momentum. Such flows are nearly spherical. Thus, the height-integrated differential equations can describe them well.
We assume all of variables are functions only of radius and the viscous accretion flow is in steady-state and axisymmetric ($ \partial/\partial t  = \partial/\partial \phi = 0 $).
The assumptions presented here are identical to 
those described by \cite{2011MNRAS.415.3721N}, except we use
the assumption of vertical hydrostatic equilibrium to compute the vertical thickness 
of the accretion disk. We adopt the cylindrical coordinate system, i.e., $ (r, \phi, z) $. Under the assumptions mentioned above, the one-dimensional vertically integrated, steady-state continuity equation read as follows

\begin{equation}\label{cont1}
\diff{}{r} (\rho r H v ) = 0,
\end{equation}

where $ \rho $ is the density of the gas, $ v $ is the radial inflow velocity, 
and $ H $  denotes the vertical scale height of the gas, respectively. Following \cite{1997ApJ...476...49N},
the hydrostatic assumption is adopted to describe the entire flow, including 
the subsonic and supersonic regions\footnote{In almost all one-dimensional 
flow models, the vertical half-thickness of the flow is approximated using the 
vertical hydrostatic equilibrium, although this assumption is not very accurate 
close to the horizon where the accretion flow becomes supersonic.}. Hence, the
flow half-thickness will be written as

\begin{equation} \label{H_r}
	H(r) = \frac{c_{\rm s}}{\Omega_{\rm K}}.
\end{equation}

In the above equation, $ \Omega_{\rm K} $ denotes the Keplerian angular velocity  
and $ c_{\rm s} \equiv (p/\rho)^{1/2} $ is the isothermal sound speed and 
$ p $ represents the gas pressure, respectively.

 In analytical works, the mass accretion rate is generally assumed to be constant. So here, with the integration of equation \ref{cont1}, the mass accretion rate $ \dot{M} $ is determined as follows 

\begin{equation}\label{mdotnew}
	\dot{M} = - 4 \pi r H \rho v,
\end{equation}

In this study, the effects of general relativity are 
modeled with the pseudo-Newtonian form of potential \cite{1980A&A....88...23P} as,

\begin{equation} \label{Phi_BH}
	\Phi_{\rm _{BH}} = - \frac{c^2}{2}\frac{r_{\rm s}}{r - r_{\rm s}}, 
\end{equation}

where $ r_{\rm s} = 2 GM / c^{2} $
is the Schwarzschild radius of the black hole of mass $ M $ ($ c $ represents the speed of light and $ G $ is the gravitational constant)\footnote{Note that
in section \ref{Phi}, the galaxy potential is also included to the total potential, $ \Phi $.}. 
The Keplerian angular velocity $ \Omega_{\rm K} $, as well as the specific angular momentum 
$ l_{\rm K} $ at the radius of $ r $, can be defined respectively by,

\begin{equation} \label{Omega_K}
	\Omega_{\rm K}^{2}(r) \equiv \frac{GM}{r \left(r - r_{\rm s} \right)^{2}}
\end{equation}
and

\begin{equation} \label{l_K}
	l_{\rm K}^{2}(r) \equiv \frac{GM r^{3}}{\left (r - r_{\rm s} \right)^{2}}.
\end{equation}

In a steady-state, the radial momentum equation can be expressed as

\begin{equation} \label{mom_r}
	v \diff{v}{r} =  r \Omega^{2} - \diff{\Phi}{r} - \frac{1}{\rho} \diff{}{r} (\rho c_{\rm s}^{2}), 
\end{equation}
where $ \Omega $ is the angular velocity of the gas on the disk mid-plane. 
We mainly focus on a condition of $ \Omega^{2} \ll \Omega_{\rm K}^{2} $,
meaning the centrifugal acceleration is extremely weak compared to the gravitational one.
In a real accretion flow, the angular momentum is transferred by Maxwell stress associated 
with magnetohydrodynamic (MHD) turbulence driven by magneto-rotational instability 
(MRI; see \cite{1991ApJ...376..214B}). In this study, we mimic the effect of MRI by adding 
viscous terms in the momentum and energy equations (see, e.g., \cite{2012ApJ...761..130Y}). 
Furthermore, the MHD numerical simulations of accretion flow show that the $ r \phi $ 
component of the magnetic stress is the most dominant component of the stress tensor
(see e.g., \cite{1995ApJ...446..741B}; \cite{1995ApJ...440..742H}). 
The kinematic viscosity coefficient is evaluated with the well-known $ \alpha $-prescription 
\cite{1973A&A....24..337S} as 

\begin{equation}  \label{nu}
	\nu = \alpha H c_{\rm s} = \alpha \frac{c_{\rm s}^{2}}{\Omega_{\rm K}},
\end{equation}

Where $ \alpha $ is a constant called Shakura–Sunyaev viscosity parameter. 
The steady-state angular momentum equation in the presence of $ r \phi $ 
component of the viscous stress tensor takes the form

\begin{equation}\label{mom_phi}
	v \diff{}{r}\left( \Omega r^{2} \right) = \frac{1}{\rho H r} \diff{}{r} \left( \nu \rho r^{3} H \diff{\Omega}{r} \right).
\end{equation}

Substituting equation (\ref{nu}) into equation (\ref{mom_phi}) and integrating of it yields 

\begin{equation} \label{domgdr}
	\diff{\Omega}{r} = \frac{v \Omega_{\rm K}}{\alpha r^{2} c_{\rm s}^{2}} \left( \Omega r^{2} - j \right)
\end{equation} 
where the integration constant $ j $ represents the specific angular momentum 
per unit mass, which is swallowed by the black hole\footnote{The quantity $ j $
is one of the two eigenvalues in this study and will be determined self-consistently 
through our physical boundary conditions}.
Eventually, the energy conservation equation can be expressed by the balance 
between the viscous heating, $ Q^{+} $, the radiative cooling, $ Q^{-} $, and the 
radial advection heat transport, $ Q^{\rm adv} $ which is expressed as

\begin{equation}
	Q^{\rm adv} = Q^{+} - Q^{-}.
\end{equation}

As mentioned, we are interested in the flow in hot accretion mode, where 
the radiative cooling term is negligibly small compared to the viscous heating ones.
Hence, we can write the energy equation of the flow in one dimension as 
 
\begin{equation} \label{energy}
	\rho v \diff{\varepsilon}{r} - c_{\rm s}^2 v \diff{\rho}{r} = \frac{ f \alpha \rho c_{\rm s}^{2} r^{2}}{\Omega_{\rm K}} \left( \diff{\Omega}{r} \right)^{2},
\end{equation}

Where the specific internal energy of the gas is given by

\begin{equation*}
\varepsilon = \frac{c_{\rm s}^2}{\gamma - 1}
\end{equation*}

And $ \gamma $ is the ratio of specific heat of the gas, set to be $ \gamma = 5/3 $. 
In the above equation, following \cite{1994ApJ...428L..13N}, the advection parameter is introduced 
as $ f = Q^{\rm adv} / Q^{+} $. Consequently, the cooling term takes the $ (1 - f) $ factor of 
the heating rate, which is an excellent approximation for inefficient radiative accretion flows 
(RIAFs). We set $ f = 1 $ throughout this paper (full advection case).

\subsection{Sonic Point Analysis}

The flow must pass a critical point (sonic radius) in transonic accretion. The critical radius needs a regularity condition that the flow must satisfy there.
Substituting equations (\ref{mdotnew}) and (\ref{domgdr}) into equation (\ref{energy}) the derivative of sound speed will be written as

\begin{multline} \label{dcsdr}
	\frac{\gamma + 1}{\gamma - 1} \diff{\ln c_{\rm s}}{r} = - \diff{\ln |v|}{r} + \diff{\ln \Omega_{\rm K}}{r} - \frac{1}{r}  \\
	+ \frac{f \Omega_{\rm K} v \left(\Omega r^{2} - j \right)^{2}}{\alpha r^{2} c_{\rm s}^{4}}.
\end{multline}

Using the above equation for $ \mathrm{d} c_{\rm s} / \mathrm{d} r $ as well as the equation (\ref{mdotnew}), 
the radial component of the momentum equation can be achieved in the form

\begin{equation}\label{dvdr}
	\diff{\ln |v|}{r} =  \frac{\mathcal{N}}{\mathcal{D}}.
\end{equation}

Here, the numerator and denominator will be given respectively as,

\begin{multline} \label{N_mathcal}
	\mathcal{N} = \left( \Omega_{\rm K}^{2} - \Omega^{2} \right) r - w_{1} \left( \frac{1}{r} - \diff{\ln \Omega_{\rm K}}{r} \right) c_{\rm s}^{2} \\
	+ \frac{w_{2} v \Omega_{\rm K} \left( \Omega r^{2} - j \right)^{2}}{r^{2} c_{\rm s}^2},
\end{multline}

\begin{equation} \label{D_mathcal}
	\mathcal{D} = C_{\rm s}^{2} - v^{2}, 
\end{equation}
with
\begin{equation}
	w_{1} = \frac{2 \gamma}{\gamma +1} \,; \quad w_{2} = \frac{f \left(\gamma - 1 \right)}{ \alpha \left(\gamma+1 \right)}.
\end{equation}

The critical radius is where the flow speed equals the characteristic speed of acoustic perturbations. This is not always the same as where the flow speed is equal to the speed of sound \cite{2008bhad.book.....K}. In equation (\ref{D_mathcal}), $ C_{\rm s} $ is the adiabatic sound speed which equals the characteristic speed of acoustic perturbations and is defined as

\begin{equation} \label{Cs}
	C_{\rm s}^{2} = w_{1} c_{\rm s}^{2}.
\end{equation}

Where $ c_s $ is the isothermal sound speed. The Mach number is given by

\begin{equation}
\mathcal{M} = \frac{\mid v \mid}{C_s}
\end{equation}

In the next section, we will explain the boundary conditions as well as the numerical technique 
for solving the system of equations.

\section{Boundary Conditions} \label{sec:BC}

In this study, we are interested in the case where the outer boundary, $ r_{\rm out} $, is at the Bondi radius, $ r_{\rm B} = GM / c_{\rm s \infty}  $, where $ c_{\rm s \infty} $ represents the sound speed at the infinity or equivalently the 
temperature of the external gas there. Here, the flow velocity is minimal compared 
to the sound speed. We name this region a subsonic zone. On the other hand, due to 
the strong gravity near the black hole, we are expecting the accretion flow becomes 
necessarily transonic, meaning that the flow velocity exceeds the 
sound speed and becomes supersonic. We call $ r_{\rm c} $ sonic point radius which is the second eigenvalue in this work and we need one extra boundary condition to determine this radius. The inner region is then called a supersonic zone. Hence, our computational domain consists of two parts. In the next subsection, we will explain these two regions' properties and boundary 
conditions.

\subsection{The Subsonic Region}

In our accretion flow model, viscosity and angular momentum play a crucial role 
in the outer subsonic region. This region starts from sonic radius $ r_{\rm c} $ to the 
outer radius, which is at the Bondi radius in this study. To find the solution in this 
region, we must solve the set of differential equations with the boundary value problem
method. Our differential equations, equation (\ref{domgdr}), (\ref{dcsdr}) and (\ref{dvdr}) 
consist of three first-order ordinary equations for three radial dependent variables 
$ v(r) $, $ \Omega(r) $ and $ c_{\rm s}(r) $ with two unknown parameter $ j $ and 
$ r_{\rm c} $. Due to the stiffness of our differential equations, we use a relaxation method 
to solve them (see, \cite{1992...}). Here, we need five boundary conditions on both ends.
The problem is that only one boundary abscissa, i.,e $ r_{\rm out} $ can be specified, 
while the other boundary $ r_{\rm c} $ is the eigenvalue and needs to be determined.
Mathematically, this case is called a {\it free boundary problem}. We reduce this problem
to the standard point by introducing a new dependent variable as 

\begin{equation} \label{y4}
	y_{4} = \ln( r_{\rm out} ) - \ln( r_{\rm c} ),
\end{equation}
where

\begin{equation} \label{dy4}
	\diff{y_4}{r} = 0.
\end{equation}

We also define a new independent variable $ t $ by setting, 

\begin{equation}
\ln(r) - \ln (r_{\rm c}) = t y_4,   \quad 0 \leq t \leq 1.
\end{equation}

Due to the presence of the new independent variable $ t $, we have two boundaries specified so that we can solve the set of differential equations with the boundary value problem method. 

\subsubsection{Critical Point Condition}

We consider three boundary conditions at the sonic point as follows: 
At $ r = r_{\rm c} $, or equivalently $ t = 0 $ the left-hand-side of 
equation (\ref{dvdr}) becomes null which represents the singularly.
To have a smooth solution crossing this point, we vanish the right-hand-side
of this equation, i.e., $ \mathrm{d} v / \mathrm{d} r = 0 / 0 $.
Therefore, we obtain two boundary conditions from equation (\ref{dvdr}) at $ r = r_{\rm c} $
as,

\begin{equation} \label{bc_Deq1}
	\mathcal{D} = 0  \Rightarrow C_{\rm sc}^{2} - v_{\rm c}^{2} = 0, 
\end{equation}
and 
\begin{multline} \label{bc_Neq1}
	\mathcal{N} = 0  \Rightarrow \left( \Omega_{\rm Kc}^{2} - \Omega_{\rm c}^{2} \right) r_{\rm c} - w_{1} 
	\left[ \frac{1}{r_{\rm c}} - \left(\diff{\ln \Omega_{\rm K}}{r} \right)_{\rm c} \right] c_{\rm sc}^{2} \\
	+ \frac{ w_{2} v_{\rm c} \Omega_{\rm Kc} \left( \Omega_{\rm c} r_{\rm c}^{2}  - j \right)^{2}}{r_{\rm c}^{2} c_{\rm sc}^2} = 0.
\end{multline} 

In the above equations, the subscript ``c'' denotes the flow quantities at the sonic point 
as well as the gradient of the radial velocity, which will be obtained by applying 
l'H\^{o}pital's rule as 

\begin{equation} \label{hopital}
	\left( \diff{v}{r} \right)_c = \left( \frac{ \mathrm{d}{\mathcal{N}} / \mathrm{d} r}
	{\mathrm{d} \mathcal{D} / \mathrm{d} r} \right)_{r = r_{\rm c}}.
\end{equation}

The angular momentum equation obtains the third boundary condition in the sonic radius. 
Following NF11, we apply the no-torque condition $ \mathrm{d}
(\Omega r^{2}) / \mathrm{d} r = 0 $ at sonic point. Since we assumed the viscosity 
vanishes at the inner supersonic region, the specific angular momentum of the 
gas becomes constant in this region, i.e., $ \Omega r^{2} = \mathrm{constant} $. The
third boundary condition at the sonic point can be simply obtained from equation (\ref{domgdr}) as

\begin{equation} \label{bc_notorque}
	\Omega_{\rm c} r_{\rm c}^{2} - j  = - \frac{2 \alpha c_{\rm sc} \Omega_{\rm c} r_{\rm c}^{2}}{v_{\rm c}}.
\end{equation}

\subsubsection{Outer Boundary Condition (OBC)}
 
The remaining two boundary conditions will be applied at the outer radius $ r_{\rm out} $.
To compare our results with NF11, we assumed that the flow begins from hot external 
media at large radii\footnote{The large radii ($ r \gg r_s $, \cite{2008bhad.book.....K}) refer to the outer edge of the accretion flow that is large enough to be near the uniform external medium. 
Furthermore, we know this is a scale of the galaxy center and is very small from the point of view of a galaxy.}. In terms of the hot accretion flow, like our galactic center, Sgr A* and M87, 
the hot gas at the Bondi radius can stay hot up to the inner edge and does not necessarily need 
any transition radius. Unlike NF11, we fix the outer boundary as $ r_{\rm out} = r_{\rm B} $.
As the first boundary condition at $ r_{\rm out} $, the values of temperature will be used there as

\begin{equation} \label{bc_cout}
	c_{\rm s \infty} = c_{\rm out} 
\end{equation}

The temperature is equivalent to the sound speed, which, in this work, we consider 
the temperature in a range of  $ \sim 10^{6}-10^7 K $ for the interstellar medium in 
the nucleus of an elliptical galaxy at the center of a cool-core group or cluster of galaxies
(see section \ref{TOUT} for more details).

The last boundary condition relates to angular velocity in the external medium. 
We assume slowly rotating gas has a constant angular velocity at the outer boundary as

\begin{equation} \label{bc_omgout}
	\Omega = \Omega_\mathrm{out}, \quad  r = r_\mathrm{out}.
\end{equation}

The angular momentum of the gas is one of the most important physical quantities 
in rotating accretion flows. Following NF11, we introduce a dimensionless 
special angular momentum. This quantity is equal to the ratio of the specific angular 
momentum at the Bondi radius $ \ell_{\rm B} $ to the specific angular momentum 
of the marginally stable orbit, $ \ell_{\rm ms} $ and is defined as,

\begin{equation} \label{eq_L}
	L = \frac{\ell_{\rm B}}{\ell_{\rm ms}}
\end{equation}

When a particle orbits circularly at $ r = r_{\rm ms} $, its specific angular momentum 
will be given by $\ell_{\rm K} (r_{\rm ms}) = \ell_{\rm ms}$. For the PW potential, $  \ell_{\rm ms} $ can be evaluated as

\begin{equation} \label{ell_ms}
\ell_{\rm ms} = \sqrt{\dfrac{27}{8}}\, c \, r_{\rm s}.
\end{equation}

The above equation can be found by substituting $ r= r_{\rm ms} $ into Equation \ref{l_K}.
When $ \ell_B $ is much smaller than $ \ell_{\rm ms} $, it is Bondi-like type accretion flow, because the Bondi accretion flow has no angular momentum ( $ \ell = 0 $ ) and when $ \ell_B $ is larger than $ \ell_{\rm ms} $, it is Disk-like type accretion flow.

  \begin{figure}
  \centering                
  \scalebox{0.5}            
  {\includegraphics{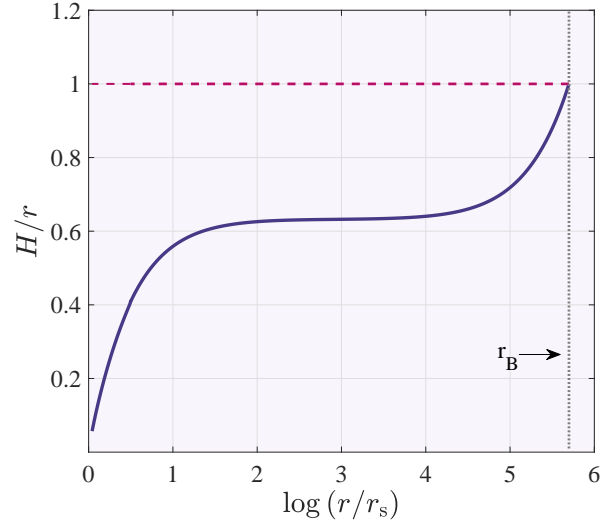}} 
  \caption{ The plot shows the radial variation of the relative thickness of the flow $ H/r $. The vertical dotted line indicates the location of the Bondi radius. }            
  \label{H3}
  \end{figure}

\begin{figure*}
\centering
\makebox[\linewidth]{%
\begin{tabular}{cc}
    \includegraphics[width=0.5\linewidth]{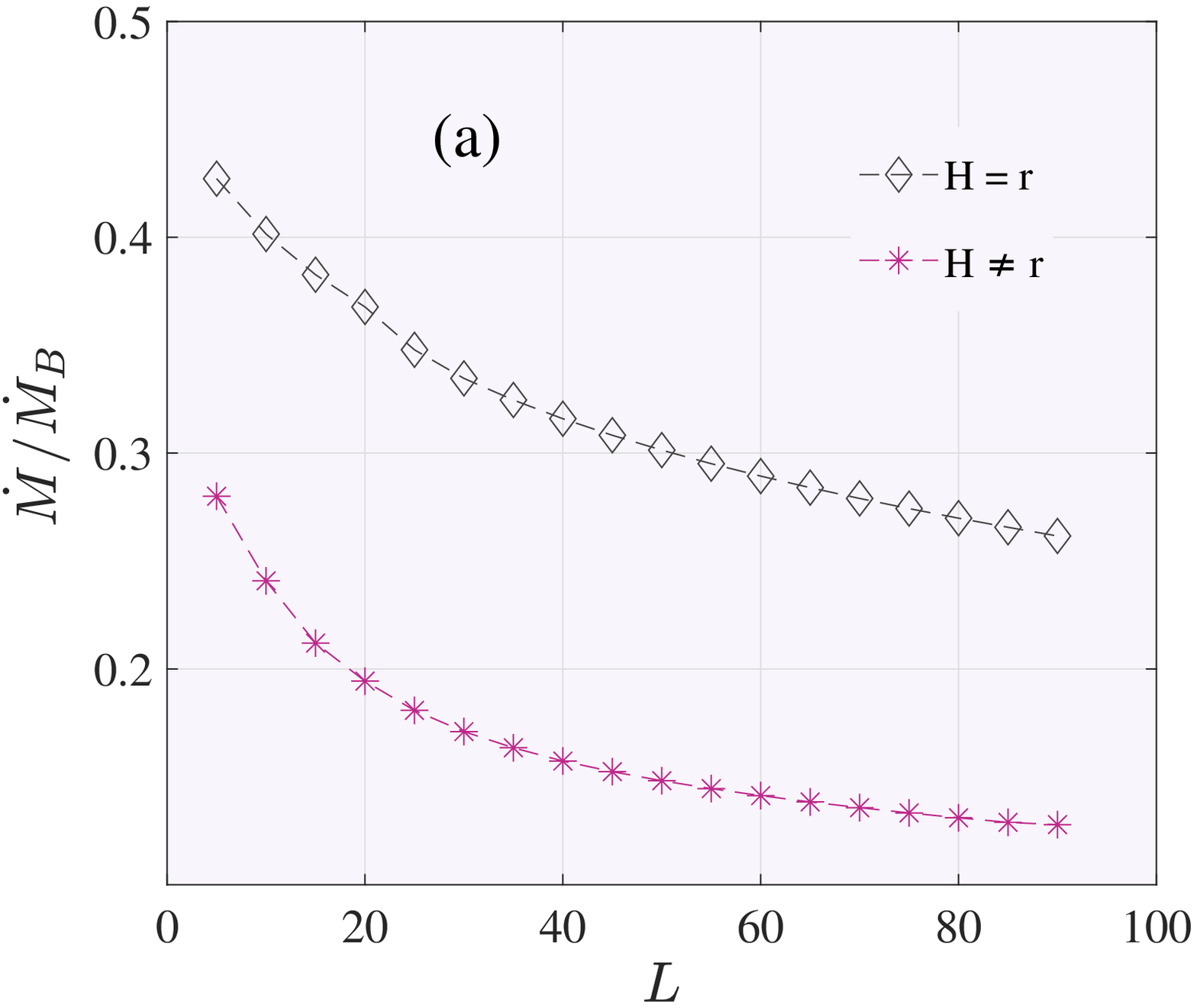} &
    \includegraphics[width=0.5\linewidth]{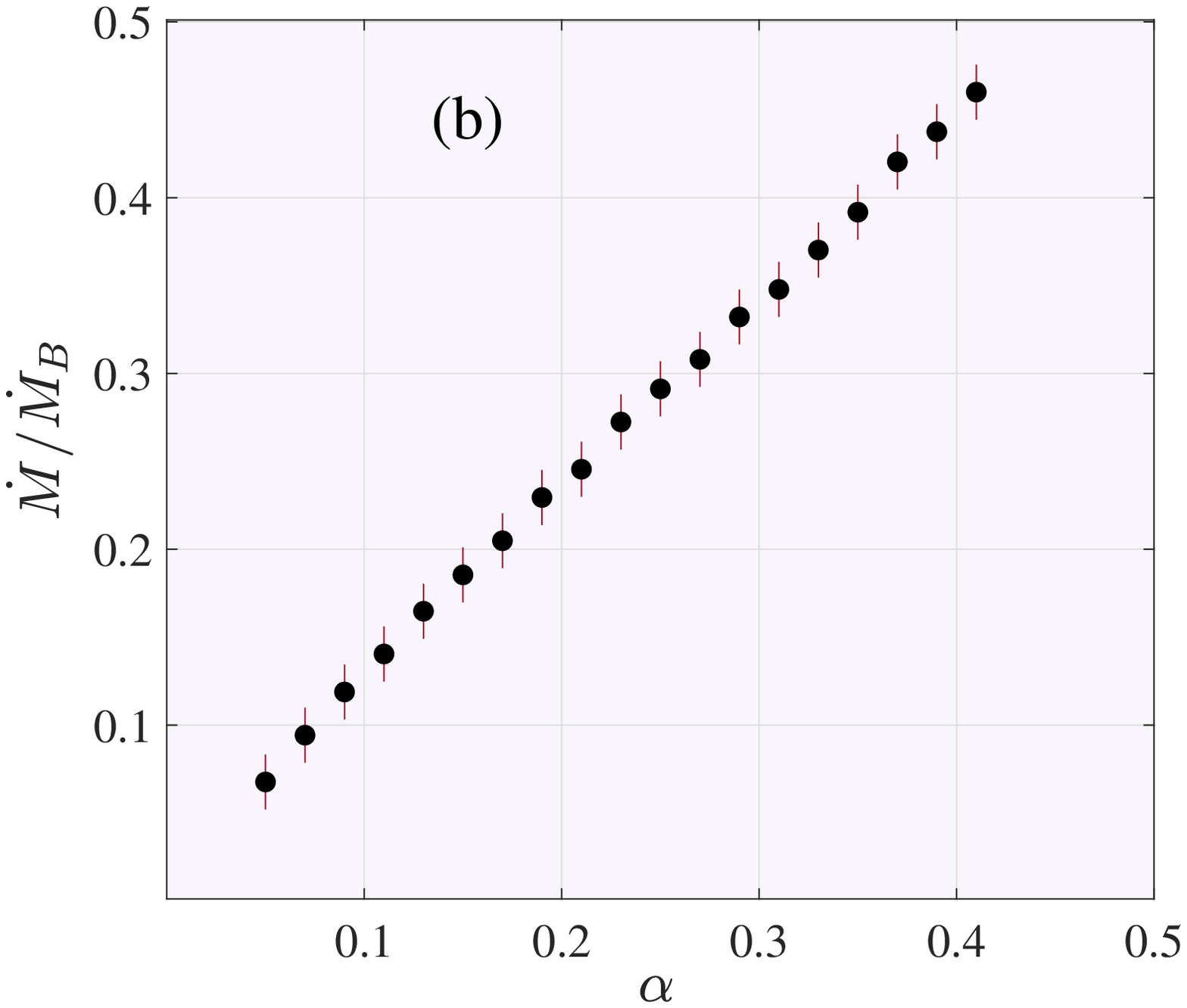} \\
    \includegraphics[width=0.5\linewidth]{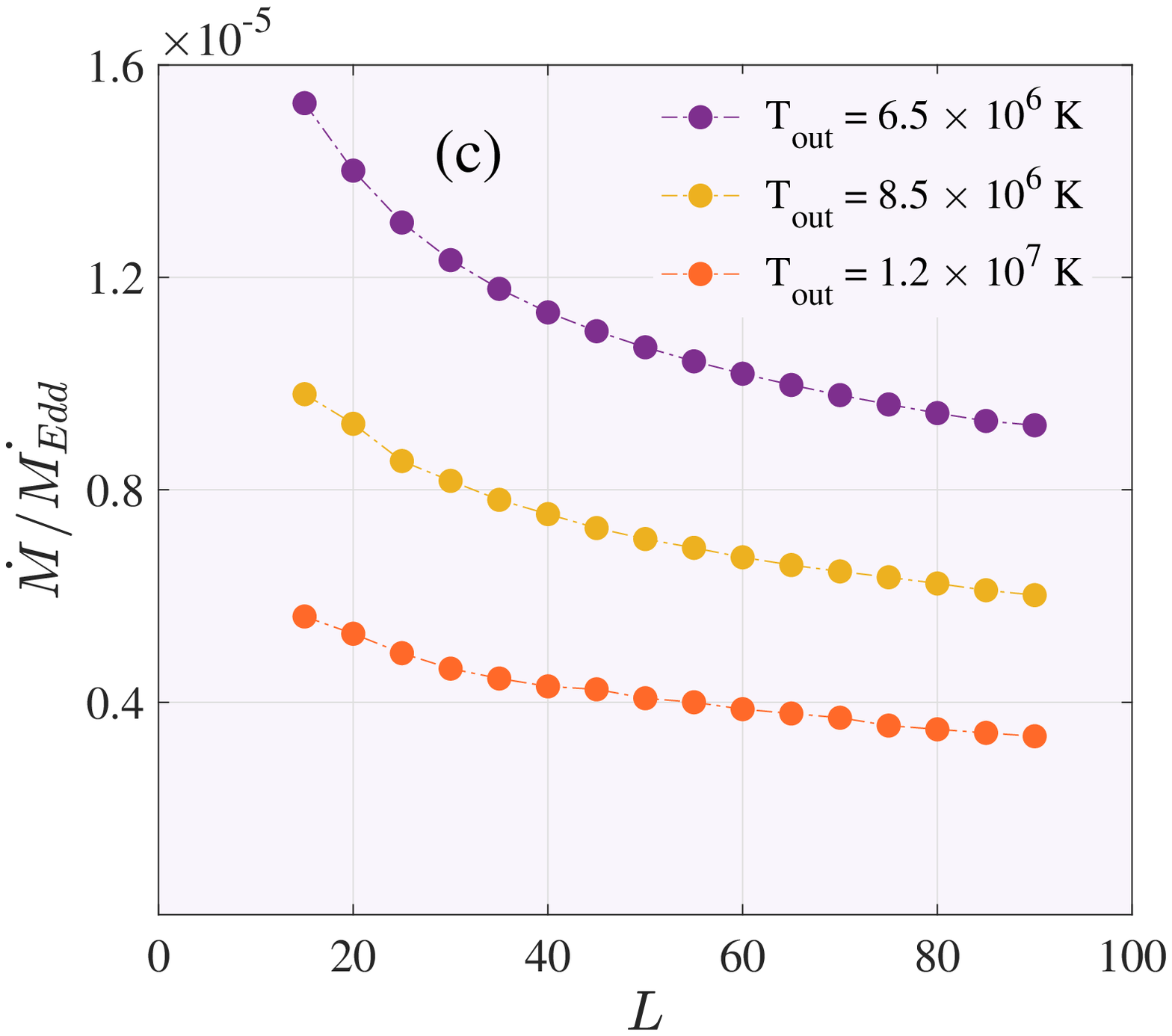} &
    \includegraphics[width=0.5\linewidth]{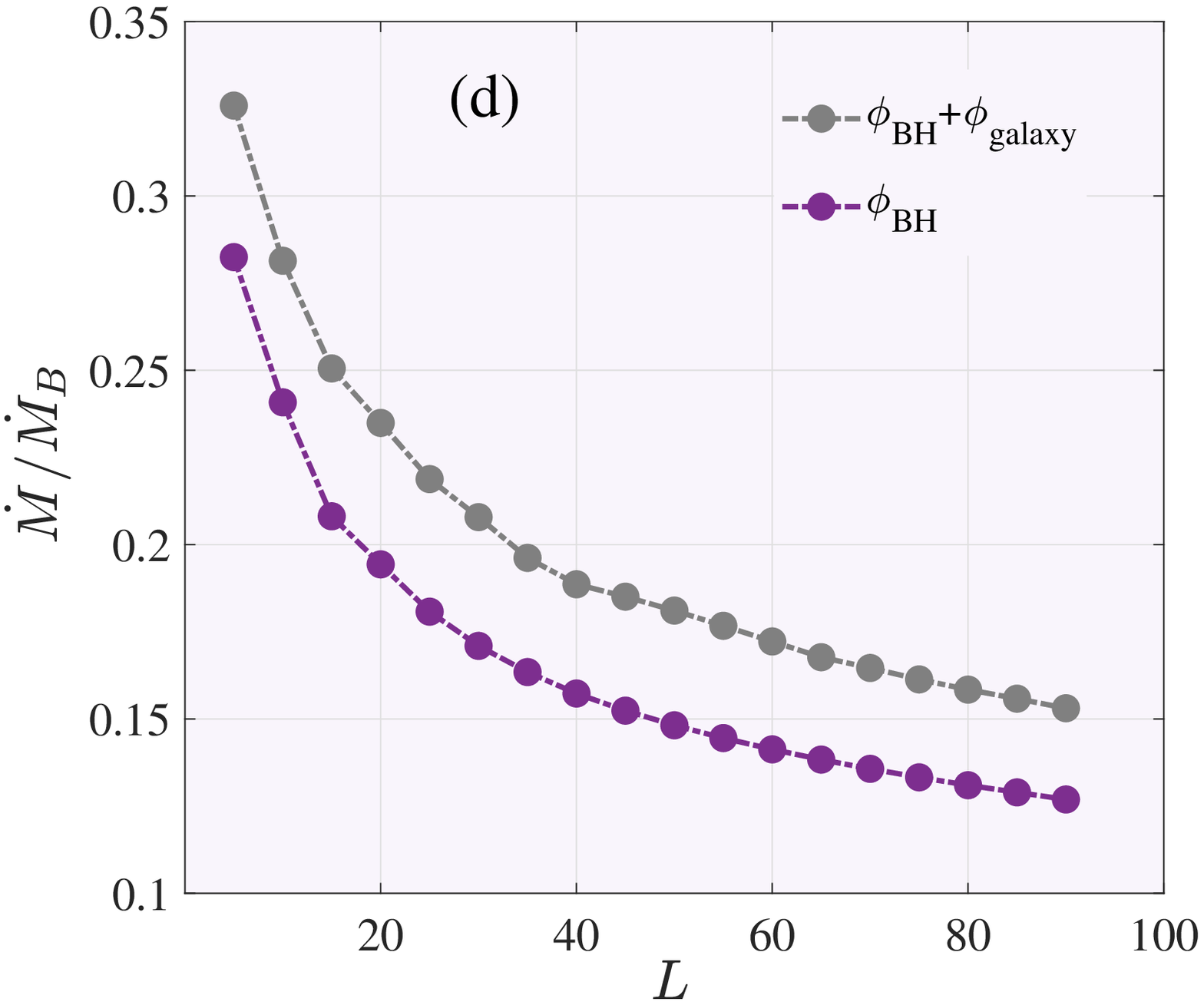}
  \end{tabular}%
  }
  \centering  
\caption{The variation of mass accretion rate as a function of physical parameters for solutions with  $ \gamma = 5/3 $. Panel (a): Mass accretion rate in unit of the Bondi accretion rate as a function of parameter $ L $, angular momentum at the outer boundary,  for solutions with $ \alpha = 0.1 $, $ T_{\rm out} = 6.5 \times 10^6 $ K. Two symbols represent the different states of the flow thickness: stars for approximate of $ H = r $ (NF11) and diamonds are for a more realistic state $ H \neq r $. 
Panel (b): Mass accretion rate in unit of the Bondi accretion rate as a function of parameter $ \alpha $  for solutions with $ L = 85 $, $ T_{\rm out} = 6.5 \times 10^6 $ K. 
Panel (c): Mass accretion rate in units of the Eddington accretion rate as a function of parameter $ L $ for solutions with $ \alpha = 0.1 $. Different colors indicate solutions for different temperatures at the outer boundary. Purple, yellow and orange colors correspond to $ T_{\rm out} = 6.5 \times 10^6 $ K,  $ T_{\rm out} = 8.5 \times 10^6 $ K and $ T_{\rm out} = 1.2 \times 10^7 $ K, respectively.
Panel (d): Mass accretion rate in unit of the Bondi accretion rate as a function of parameter $ L $ for solutions with $ \alpha = 0.1 $ and $ T_{\rm out} = 6.5 \times 10^6 $. Solutions without considering galaxy potential are represented by purple dots, while solutions with galaxy potential are represented by gray dots (For more information, see section \ref{Phi}). }
\label{11} 
\end{figure*}

\subsection{The Supersonic Region} \label{sec:inner_region}

As we mentioned before, the motion of the flow near the BH is physically significant. 
When the matter passes the sonic point, $ r = r_{\rm c} $, the accretion flow tends 
to be supersonic. The transition from a subsonic flow to a supersonic occurs in a narrow 
region in the radial direction, and viscosity decreases in this region. 
Following NF11, we assume the viscosity vanishes in this region
(see \cite{1992ApJ...394..261N}, \cite{1994PASJ...46..289K}; \cite{1997MNRAS.285..239K} for more details).
Thus, we drop all terms corresponding to the viscosity from our differential 
equations and set $ \alpha = 0 $.  
By ignoring $ \alpha $ from the differential equations, the equations simplify to become 
a set of algebraic equations. Consequently, the specific angular momentum, $ \ell_{\rm in} $, 
entropy, $ \rm{S}_{\rm in} $ and Bernoulli parameter, $ \mathcal{B} $ are constant in this zone
and evaluated as:

\begin{equation}
	\ell_{\rm in} \equiv \Omega r^{2} = \mathrm{constant},
\end{equation} 

\begin{equation}
	\mathrm{S}_{\rm in} \equiv \frac{c_s^2}{\rho^{\gamma - 1}} = \mathrm{constant},
\end{equation} 
and 
\begin{multline}
	\mathcal{B} \equiv\frac{v^2}{2} + \frac{\ell_{\rm in}^2}{2 r^2} + \phi + \frac{\gamma \mathrm{S}_{\rm in}}{(\gamma - 1)} (\frac{\dot{M} \Omega_k}{4 \pi r  \mathrm{S}_{\rm in}^{1/2} |v|})^{2\frac{(\gamma-1)}{(\gamma+1)}} \\  = \mathrm{constant}.
\end{multline}

We first compute the value of $ \ell_{\rm in} $, $ \mathrm{S}_{\rm in} $,
$ \mathcal{B} $, and $ \dot{M} $ at the sonic point, $ r = r_{\rm c} $ and then
directly evaluate the solution from the above algebraic equations for the inner
supersonic region. It is worth mentioning that in terms of MHD case, studying 
the inner region of the accretion flow close to the central black hole is important for 
understanding the jet ejection and also in the magnetically arrested disk, MAD 
(see e.g., \cite{1976Ap&SS..42..401B}).

\section{NUMERICAL RESULTS}\label{3}

This paper aims to obtain numerical exact global solutions. The global solutions contain flow properties near the inner or outer boundary, while the self-similar solutions cannot describe flow in the region close to the boundaries (\cite{1994ApJ...428L..13N}).

In hot accretion flows, it is common that a thick disk approximation ( $ H = r $ ) is often taken for simplicity. Unlike other studies, we stick to maintaining hydrostatic equilibrium along with vertical directions. This means that we take the initial state into account ($ H \neq r $) because it is more realistic. The present study will also lead to understanding how the flow thickness changes in the direction of radius. Figure \ref{H3} shows the variation of the relative half-thickness of the flow for all the ranges of radii. It is seen that the flow is thin in its inner region and is quasi-spherical ( $ H/r $ is close to 1) in the outer region (Near the Bondi radius). It is notable that at intermediate radii including the radial range of ($ 10 $ - $10^5$) $ r_s $, the flow thickness is approximately constant ( $ H/r \sim 0.6 $ ). Since $ H/r $ is the ratio of pressure force to centrifugal force (or gravitational force), we expect in the vicinity of the black hole ($ r \sim 10$ $ r_s $), due to the strong gravity of the black hole, the $ H/r $ tends to zero. Previous studies have taken the outer boundary condition as a thin disk, which causes the thickness of the flow to decrease rapidly with a radius beyond the outer boundary (\cite{1999ApJ...523..340L}, NKH97). However, this study assumes that the gas starts in a state corresponding to a real interstellar medium, not a thin accretion disk.

\begin{figure*}
\centering
\includegraphics[width=85mm]{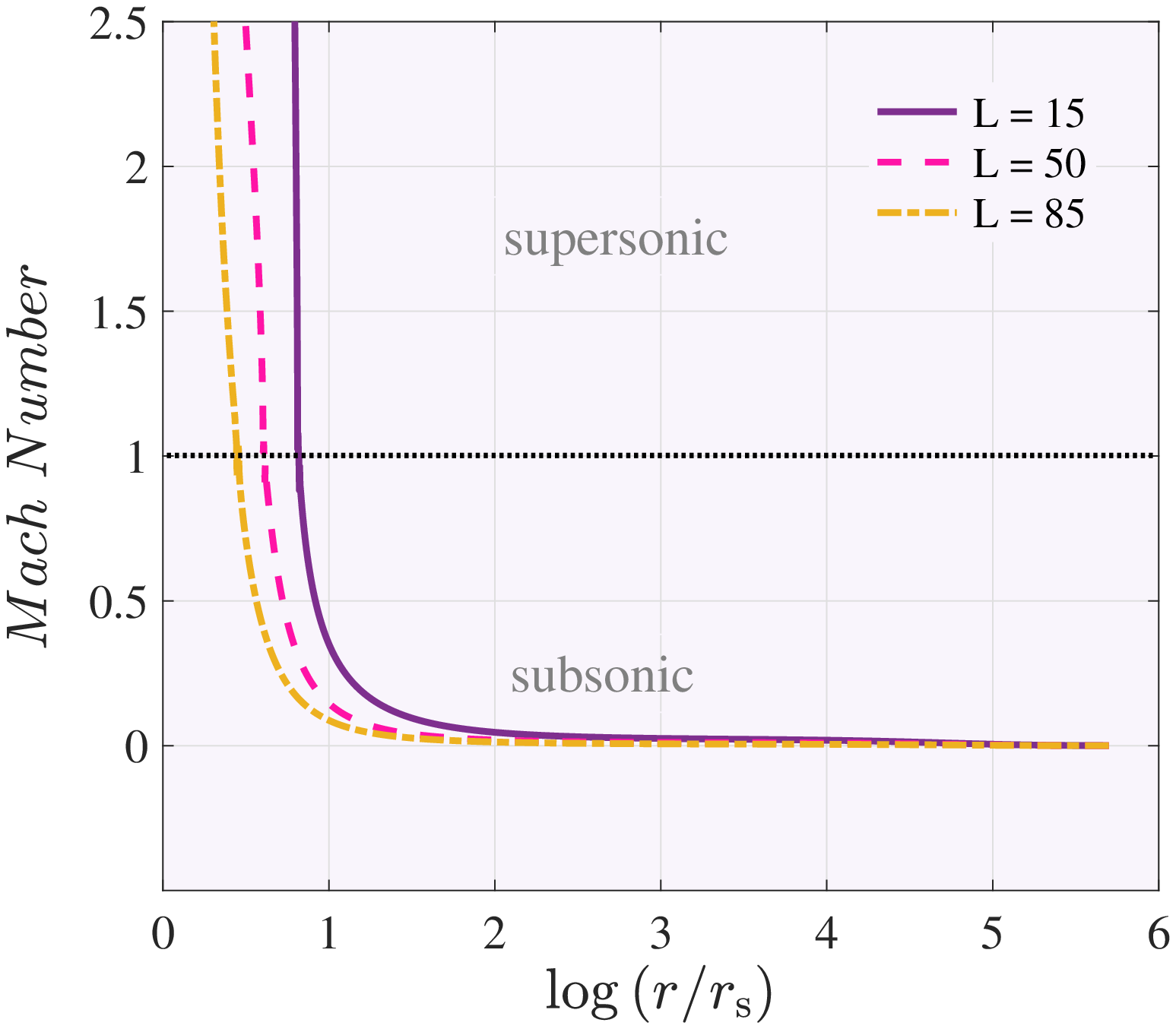}
\includegraphics[width=85mm]{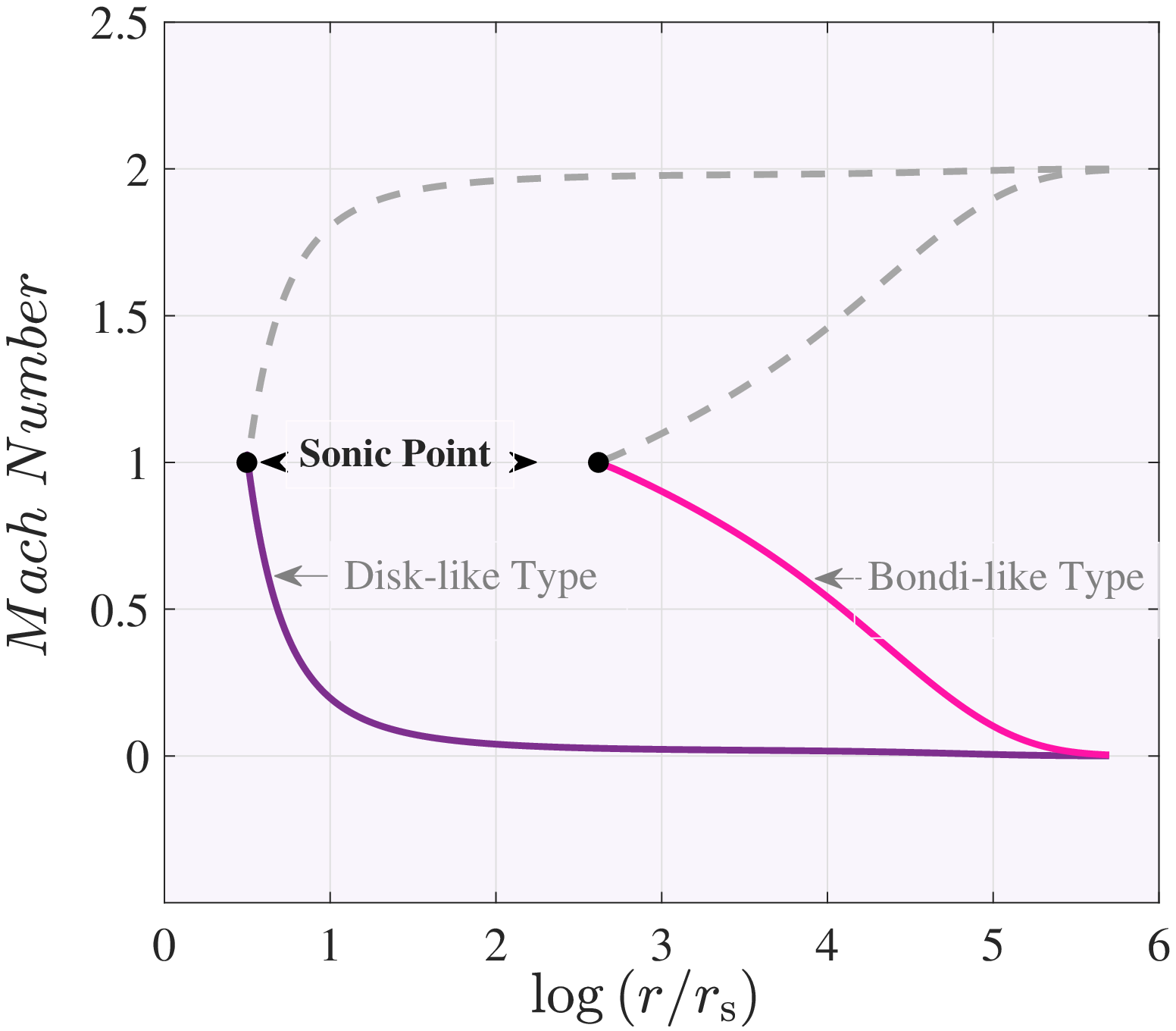} 
\caption{Global numerical solutions for the Mach Number. Left: Comparison of solutions (Disk-like Type) for different $ L $. The solid,  dash and dash-dot lines are for $( L, r_c ) = (15, 6.483 ); ( 50, 4.017 ); (85, 2.787 )$  respectively and $\alpha = 0.05$. Right: Solutions include the subsonic region which is for  $( L, r_c ) = ( 85, 3.144 ) $ (Disk-like Type) and  $ ( L, r_c ) = ( 0.11, 415.446 ) $ (Bondi-like Type) and $\alpha = 0.1$. The numerical solutions are between the outer boundary and the critical point.}

\label{12} 
\end{figure*}
 
\subsection{Effect of the Outer Boundary Conditions}\label{TOUT}
In this section, we investigate the effects of the outer boundary condition (OBC) on the dynamics and the structure of slowly rotating accretion flows. We find that OBC plays an important role. In many accretion flows, external boundary conditions like angular momentum, temperature, and density significantly influence flow properties, such as velocity and the mass accretion rate.  
In addition, the complex astrophysical environments such as interstellar mediums in the nuclei of galaxies, create different modes for the accreting gas at the outer boundary, like the temperature and angular momentum (\cite{1999ApJ...521L..55Y}, \cite{2001AIPC..556...93Y}).
So investigating the role of the OBC in accretion flows around the black holes is particularly important due to the complexity of astrophysical environments. 
 
In the following, we focus on the mass accretion rate of the accretion flow, mainly its dependence on the angular momentum of gas at the outer boundary. The dependence of the mass accretion rate on boundary conditions can only be addressed by constructing global solutions.

In accretion systems, the most important parameter is the mass accretion rate, which the gas supply environment can determine (e.g., gas temperature, density, and angular momentum), directly influencing the boundary conditions. One of the essential scales which is being utilized the accretion rate is the Eddington accretion rate $ \dot{M}_E = 10 L_E / c^2 = 1.39 \times 10^{18} (M/M_{\odot}) g s^{-1}  $. If the mass accretion rate is less than $ 2 \%$  of $ \dot{M}_E $, we have hot accretion flow (\cite{1994ApJ...428L..13N}). For $ 2 \% \dot{M}_E  < \dot{M} < \dot{M}_E $ the mass accretion rate consistent with a thin accretion disk. When the mass accretion rate is in the range $ \dot{M} \gg \dot{M}_E $, it consistent with thick accretion disk and for slim accretion disk models, accretion rate is $ \dot{M} \approx \dot{M}_E $ (\cite{1988ApJ...332..646A}).The other important scale is the mass accretion rate in spherical accretion, known as the Bondi accretion rate. \cite{1952MNRAS.112..195B} found that the mass accretion rate for the Bondi model is directly determined by the boundary conditions far from the accreting object, including temperature and density of surrounding gas. The Bondi rate is

\begin{equation*}
\dot{M}_B = \Lambda \gamma^{-3/2} 4 \pi r_B^2 \rho_{\rm{out}} c_{\rm{out}}
\end{equation*}

The constant $\Lambda(\gamma)$ is $ 0.25 $ for $\gamma = 5/3$. The Bondi radius is as $ r_B = GM/c_{\rm out}^2$. In the Bondi radius, gravity overcomes the gas pressure and begins to fall onto the black hole. $\dot{M}_B$ in our unit ($ r_g = c = \rho_{\rm out} = 1$) is equal to $3.65 \times 10^8$, while $ T_{\rm out} = 6.5 \times 10^6$K. Both units of mass accretion rate ($ \dot{M}_E $ and $\dot{M}_B$) have applications in the present study.

\begin{figure*}
\centering
\includegraphics[width=85mm]{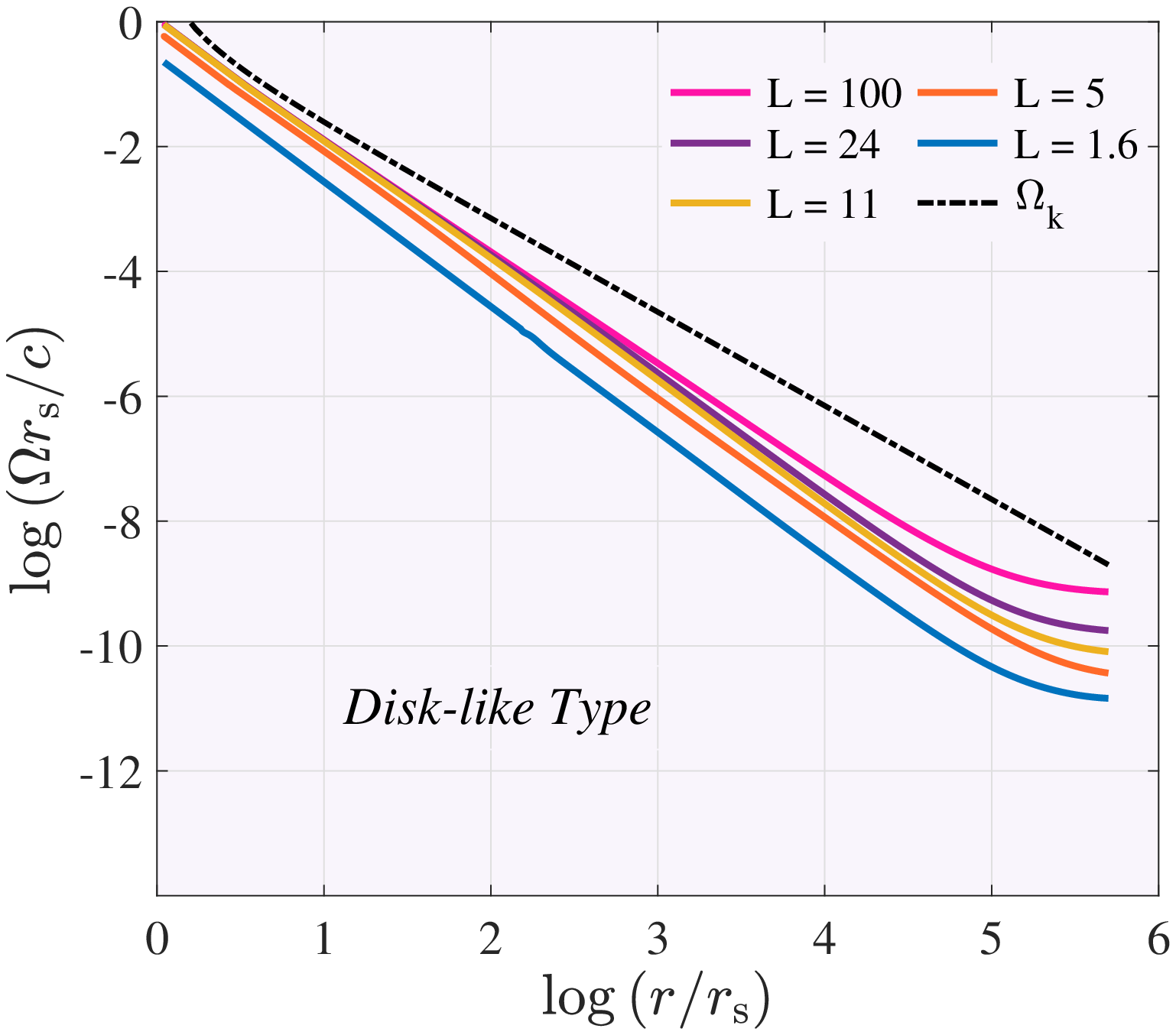} 
\includegraphics[width=85mm]{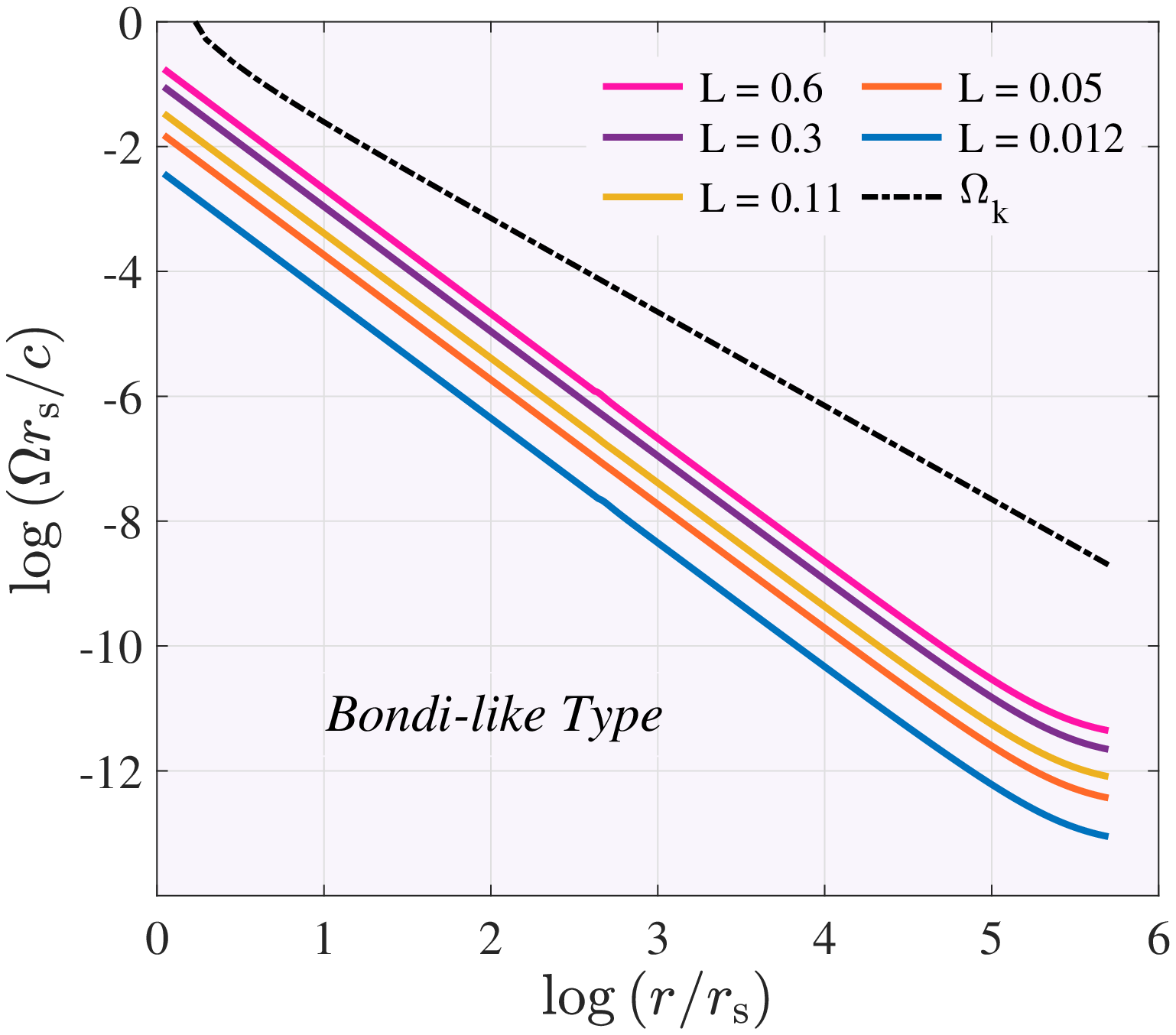} 
 \caption{ Radial variations of the angular velocity of the slightly rotating accretion flows with different parameters of $ L $. Left: The solid lines are solutions of angular velocity for five values of  $ L: 100, 24, 11, 5, 1.6 $ (Disk-like Type). The dot-dashed line shows the Keplerian angular velocity $ \Omega_k $. Right: Solutions of angular velocity are for five value of $ L : 0.6, 0.3, 0.11, 0.05, 0.012 $ (Bondi-like Type). Other parameters are $ \alpha = 0.1 $, $ \gamma = 5/3 $, $ T_{\rm{out}}= 6.5 \times 10^6 $ K.}
\label{BD_like}
\end{figure*}

Our results show that the mass accretion rate in slowly rotating accretion flow is lower than the Bondi rate, which is compatible with NF11 results. In the present study, we consider the mass accretion rate as a constant parameter at every radius; in other words, our model has no mass outflow. The mass accretion rate depends on angular momentum at the outer boundary and for $ \ell_B > \ell_{ms} $ it value is smaller than the Bondi accretion rate and for $ \ell_B < \ell_{ms}$  approximately equal with $\dot{M}_B$. 

In Figure \ref{11}, the mass accretion rate is shown as a function of angular momentum at the outer boundary (Panels (a), (c), (d)). It is obvious that when the angular momentum flow is low, the mass accretion rate is very close to the Bondi rate, but when the angular momentum flow is high, the rate is lower than the corresponding Bondi rate.
\cite{2011MNRAS.415.3721N} have used a simple approximation for the thickness of flow. They have assumed that flow geometrically is thick and $ H = r $. In this work, we have set $ H \neq r$, and our results are compared to NF11 in panel (a) of Figure \ref{11}. Surprisingly this correction leads to a change in the value of the mass accretion rate. Figure \ref{11}, panel (a) shows the variation of mass accretion rate as a function of parameter $ L $, angular momentum at the outer boundary. As we can see from panel (a), the mass accretion rate decreases significantly when we consider a more realistic mode ($ H \neq r $) instead of a geometrically thick disk ($ H = r $). Indeed keeping the assumption of the vertical hydrostatic equilibrium causes the mass accretion rate to be reduced. In Panel (b) of Figure \ref{11}, the mass accretion rate is shown as a function of viscosity $ \alpha $. Our solutions contain 19 different $ \alpha $ in the range $ 0.01 \leq \alpha \leq 0.45 $ and $ L > 1 $ (Disk-like types). As it is obvious, there is a linear relationship between the mass accretion rate and parameter $\alpha$. This linear relation can also be seen in the case of ADAFs (\cite{1994ApJ...428L..13N}, NF11). Increasing $ \alpha $ helps to accretion process and increases the mass accretion rate onto the black hole. This case study well confirms the importance of viscosity in accretion flows.  
In the following, we plan to investigate the effects of temperature at the outer boundary on the structure of slowly rotating accretion flows.

A halo of hot gas surrounds most large elliptical galaxies, and its temperature range is 1-10 million Kelvin. (It is in the range of the virial temperature) (\cite{1999ASPC..163..119K}, \cite{2014PhRvD..90d4010R}). In this study, we assume that gas accretes from the edge of a hot gas located in the Bondi radius. Beyond the Bondi radius, there is an interstellar medium, which is in the range of $10^6-10^7$ K in real position. Our focus is on how the outer boundary gas temperature affects the accretion flow structure. 
Figure \ref{11}, panel (c) illustrates the variation of mass accretion rate $ \dot{M} $ with the angular momentum of the outer boundary for different temperatures at the outer boundary. We plot the mass accretion rate in a unit of $\dot{M}_E$ versus parameter $ L $. Since the Bondi accretion rate is highly dependent on the temperature of the embedded cloud, $ T_{\rm out} $, it seems more reasonable to use the Eddington accretion rate, which under any circumstance is constant. Panel (c) in Figure \ref{11} shows that by increasing the temperature of the outer boundary, the mass accretion rate slightly decreases. We consider three different values for temperature at the outer boundary as:  $ T_{\rm out} = 6.5 \times 10^6  $ K, $ T_{\rm out} = 8.5 \times 10^6  $ K and $ T_{\rm out} = 1.2 \times 10^7  $ K. We find that regardless of $ T_{\rm out} $, the mass accretion rate $ \dot{M} $ perceptibly decreases as the angular momentum $ L $ increases. It is clear that for a given $ L $, the mass accretion rate can be very different with the changing temperature at the outer boundary. This result has also been reported in simulations (BY19). Panel (d) of Figure \ref{11} shows the effect of the potential of the galaxy on the mass accretion rate. A very slight change in the mass accretion rate is observed when galaxy potential is included in our solutions (see section \ref{Phi} for more detail).

  \begin{table*}
  
 \label{tab:natbib}
 \begin{tabular}{llllll}
  \hline
  $ L $ & $T_{out}$ & $\phi_{galaxy}$ & $ r_c $ & $ j $ & $ \dot{M}$\\
   & ($K$) &  & ($ r_s $) & ($r_s c$)  & ($ \dot{M}_B$)\\
  \hline
   & Disk-like, &  ($H \neq r $)  \\
  \hline
  $ 85 $ & $6.5 \times 10^6$   & OFF & 3.1447 & 1.0245 & 0.1289\\
  $ 85 $  & $6.5 \times 10^6$  & ON  & 3.2581 & 1.0204 & 0.1558\\
  $ 15 $ &  $6.5 \times 10^6$  & OFF & 3.2421 & 1.0004 & 0.1805 \\
  $ 15 $ &  $6.5 \times 10^6$  & ON  & 3.3789 & 0.9792 & 0.2505\\
 \hline
 & Bondi-like,  & ($H = r $)   \\
 \hline
 $  0.1 $ &  $ 6.5 \times 10^6 $ & OFF  & 415.95 &  0.026 & 0.543 \\
 $  0.1 $ &  $ 6.5 \times 10^6 $ & ON   & 441.69 &  0.023 & 0.595\\
  \hline
 \end{tabular}
 \centering
 \caption{Effect of boundary conditions on Sonic point, $ j $ and Mass accretion rate. Cols 1 and 2: The angular momentum and temperature at the outer boundary, respectively. Col 3: the galaxy potential is included ``ON'' or not ``OFF''. Col 4: the position of the sonic point. Col 5: an integration constant with dimensions of specific angular momentum swallowed by the black hole. Col 6: the mass accretion rate in a unit, the Bondi accretion rate.}
  
  \label{table}
\end{table*}

The angular momentum at the outer boundary affects the structure of the flow. Here we tend to focus on the influence of angular momentum $ L $ on the accretion of gas onto a black hole. The main result is to find two separate regimes, quasi-spherical accretion for small $ L $ and Disk-like accretion for large $ L $, respectively. There is no smooth transition between the two types of flows. For the first time, (\cite{1981ApJ...246..314A}) found a discontinuity in their study of the adiabatic accretion flow with a constant specific angular momentum between Disk-like and Bondi-like types. In this study, concerning the previous works, we corroborated that this transition exists even though the viscosity of flow is considered. The following brief description is of the two types.

{\tt Disk-like type :}
In this flow type, $\ell_B  >  \ell_{ms}$ the sonic point occurs roughly in small radii. When the specific angular momentum is high, the centrifugal force plays a dominant role rather than the gravitational force. The gas becomes supersonic only after passing through a sonic point near the horizon. {\tt Bondi-like type :} When $\ell_B  <  \ell_{ms}$, accretion flow is quasi-spherical accretion. In such a case, centrifugal force is too weak to gravity force, so the flow can not be in hydrodynamical equilibrium.
This accretion flow is similar to spherical accretion when the flow has zero angular momentum. Because of this, \cite{1981ApJ...246..314A} named it Bondi-like accretion. 
We prefer to substitute $ H \simeq r $ in conservation equations of this type. The low specific angular momentum of the accretion gas (Bondi-like type) is a new state, distinguished by its large sonic radius. 

Figure \ref{12} displays the radial variations of the Mach number with different specific angular momentum at the outer
boundary $ r_B $. At the sonic point, the Mach number equals 1. The right panel shows the Mach number for the two types of accretion flow, which corresponds to the relatively low and high angular momentum at the outer boundary, respectively. Solutions are presented for $ L = 85 $ (Disk-like type) and $ L = 0.11 $ (Bondi-like type) for $\alpha = 0.1$ that their sonic radii are (purple line) $ \sim 3 r_s $  and (pink line)$ \sim 415 r_s $, respectively. Solutions for the subsonic region are included in this panel. The left panel shows variation of Mach number for three values of $ L $ when $\alpha = 0.05$. For $ L = 15$ , location of sonic point is larger than $ L = 85$. Indeed, the sonic point moves to a larger radius as $ L $ decreases. In this panel, solutions include both subsonic and supersonic regions.
Figure \ref{BD_like} illustrates the global solutions of slowly rotating accretion flows for the different parameters of $ L $. Solutions are presented for two types of the accretion flow, the left panel according to $ L \geq 1 $ ($ L = 100, 24, 11, 5, 1.6 $), which is equivalent to a Disk-type accretion flow, and the right panel according to $ L < 1$ ($ L = 0.6, 0.3, 0.11, 0.05, 0.012$), which is equivalent to a Bondi-type accretion flow. Solutions clearly illustrate how angular momentum at the outer boundary affects the angular velocity. This figure shows that the angular velocity decreases as the angular momentum changes at the outer boundary for every two types of accretion. To a better understanding of the difference between the angular velocity of our solutions and the Keplerian angular velocity, we plot the dot-dash line that corresponds to the Keplerian angular velocity, which shows our solutions are sub-keplerian. This result emphasizes the importance of considering the outer boundary conditions of the accretion flow. 

As a final point, we must note that the effect of OBCs is subject to the location of the outer boundary. When $ r_ {\rm out} $ is large, $ T_ {\rm out} $ has a limited range, therefore its effect is minimized. When applying one-temperature plasmas, the effect of variations in $ T_{\rm out} $ on the structure of flows is reduced. However, the effect of $ \Omega_{\rm out} $ is always evident. Typically, when $ r_{\rm out} $ is small, the effect of OBCs is most apparent (\cite{2001AIPC..556...93Y}).

\subsection{Effect of the galaxy potential}\label{Phi}

Recent studies have examined numerous approaches to model a realistic approximation for the dynamics of black hole accretion flows. One of the works involves considering the gravitational potential of host galaxies. For example, the study carried out by \cite{2022MNRAS.512.2474M} incorporated the effects of the gravitational field of the host galaxy on the flow of Bondi accretion. In this section, we consider the contribution of the gravitational potential of both the central black hole and the stars of the host galaxy. At a larger scale (near $\sim$ 10 parsecs), in addition to the potential of the BH, the gravitational potential of stars of the host galaxy will become essential and should be included (\cite{2018MNRAS.478.2887Y}). This is while previous studies neglected the gravitational force of stars at parsec scale NF11, NKH97. So the total gravitational potential is given by

\begin{equation}
\phi = \phi_{BH} + \phi_{Galaxy}
\end{equation}

 The potential of the stars of the host galaxy is 
 
 \begin{equation}
 \phi_{Galaxy} = \sigma^2 ln r + C
 \end{equation}
 
  \begin{figure*}
\centering
    \includegraphics[width=120mm]{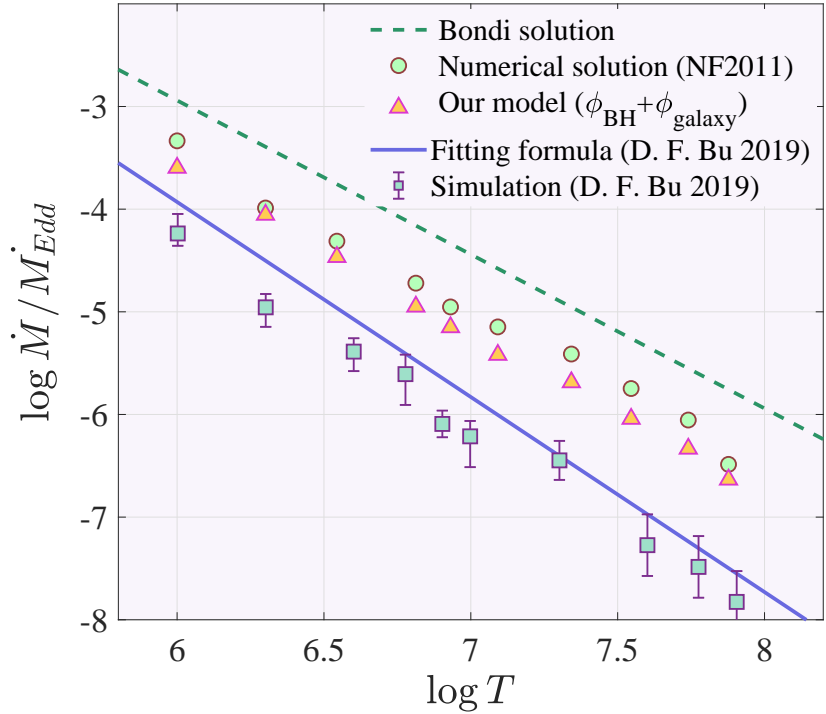}\\
\caption{ Black hole mass accretion rate in a unit of $\dot{M}_E$ as a function of the temperature at outer boundary $T_{\rm out}$. The density at outer boundary  is considered $\rho_0 = 10^{-24} $ $gr/cm^3$. Solution is for $ L = 85$, $ \alpha = 0.1 $ and $ \gamma = 5/3 $. Solid lines depict the fitting formula of BY19. By assuming $ \gamma = 5/3 $, the dashed line corresponds to the Bondi formula. The green squares represent the mass accretion rate time-averaged of simulations in BY19. Error bars overlaid on squares represent the change range of simulations due to fluctuations. The orange triangles correspond to the mass accretion rates calculated by our numerical solutions taking into account the galaxy potential, and the green circle shows the value of the mass accretion rate calculated by the model of NF11 (without the effect of galaxy potential).} 
\label{simulation} 
\end{figure*}
 
Where $\sigma$ is the velocity dispersion of stars and $ C $ is a constant (\cite{2016ApJ...818...83B}). The term $  \phi_{Galaxy} $ will add to radial momentum equation \ref{mom_r} and can effect on dynamics of flow. In elliptical galaxies with central black hole to mass $ M_{BH} = 10^8 M_{\odot}$ is common the velocity dispersion be about 150 - 250 $ km $ $ s^{-1}$. Here, we assume the velocity dispersion of stars is a constant of radius, which is  $\sigma = 200 $ $ km $ $ s^{-1}$ (\cite{2013ARA&A..51..511K}, \cite{2006ApJ...641L..21G}). The stellar velocity dispersion has a crucial contribution to the gravitational force distribution of the galaxy ( \cite{2018MNRAS.478.2887Y}). It has been found in previous studies (\cite{2016ApJ...818...83B}, \cite{2018MNRAS.478.2887Y}, \cite{2021MNRAS.505.4129S}) that the gravitational effects of stars of the galaxy can become significant when the radius is beyond $ 1 $ pc. Because of this, in our study, the effect of gravity of stars is ignored close to the black hole (the supersonic region $r\leq 1$pc). Therefore, the algebraic equations in Section \ref{sec:inner_region} do not change. Figure \ref{11}, Panel (d) illustrates variations of the mass accretion rate ($ \dot{M}/\dot{M}_{B} $) as a function of the angular momentum at the outer boundary ($ L $). There is a clear difference when we turn on the effect of galaxy potential.  The star's gravity can slightly enhance the mass accretion rate onto the black hole. This finding is in agreement with simulations of YB19 and \cite{2018MNRAS.478.2887Y}. As shown in Figure \ref{11}, Panel (d), the gravitational potential of the galaxy has a greater effect on the mass accretion rate for $ L \leq 20 $ (roughly Bondi-like type). This could be one of the reasons most recent Bondi solutions are taking into account the galaxy's potential alongside the black hole's potential. (\cite{2022MNRAS.512.2474M}, \cite{2019MNRAS.489.3870S}, \cite{2017ApJ...848...29C}). Our numerical solutions with or without nuclear stars' gravity are summarized in Tables \ref{table}. In column 1 of Tables \ref{table}, we show the values of angular momentum at the outer boundary (dimensionless). It should be noted that in this paper, we study low angular momentum accretion.  In Table \ref{table}, we have two types of accretion, Disk-like, and Bondi-like. For both states, the temperature at the outer boundary is $ T_{\rm out} = 6.5 \times 10^6 $ K. In Table\ref{table}, the temperature and gas properties in our model are the same as those in the NF11 model. The values of mass accretion rate of numerical solutions are listed in column 6 of Tables \ref{table}. It appears that the galaxy potential does significantly influence the mass accretion rate. Considering the star gravity slightly increases the mass accretion rate in our model. Our results in columns 4 and 5 indicate that the values of $ r_c $ and $ j $ do not rely much on galaxy potential and only slightly increase the radius of the sonic point. We also observe that by maintaining hydrostatic equilibrium ($ H \neq r $) within our model, the eigenvalues such as the sonic radius and $ j $ do not change significantly. When we compare our results with those of \cite{2011MNRAS.415.3721N}, we see that there is only a significant change in the value of the mass accretion rate.

\subsection{Comparison with previous works}
\cite{2019MNRAS.484.1724B} performed the two-dimensional simulations to investigate slowly rotating accretion flows irradiated by an LLAGN at parsec scales. They found the fitting formula based on temperature and density at the parsec scale, which could predict the luminosity of observed low-luminosity AGNs. As \cite{2019MNRAS.484.1724B} investigated accretion flow with low angular momentum and considered the potential of the galaxy in their study, we can compare our results with theirs. To compare this work with our model, we assume the black hole mass equal to $ 10^8 M_{\odot} $ ($ M_{\odot} $ is solar mass) and the density at the outer boundary  $\rho_0 = 10^{-24} $ $ g$ $ cm^{-3}$.

Figure \ref{simulation} compares the results of our numerical solutions and previous studies (numerical solution NF11, simulation YB19), which shows the variations of black hole accretion rate as a function of temperature at the outer boundary. In our study, the trend in the mass accretion rates versus temperature is similar to YB19. It is clear that for a given constant gas density at the outer boundary, the mass accretion rate will become smaller by increasing of temperature at the outer boundary. There is a significant difference between our results with simulation because we do not consider the effect of outflow and radiation in our study. In this Figure, the green circles correspond to the values of the mass accretion rate of solutions derived from the model of \cite{2011MNRAS.415.3721N} and the orange triangle represents our numerical solutions by considering the effect of galaxy potential. The green squares are according to the time-averaged values of the mass accretion rate of simulations in BY19, and the error bars on squares correspond to the change range of simulations due to fluctuations. In comparison to YB19 simulations, our results deviate slightly since the effect of radiation and outflow has been ignored. Hence all the gas captured at the outer boundary can fall into the black hole. The dashed line shows the Bondi accretion rate by assuming $\gamma = 5/3$. The solid line shows the black hole accretion rate according to the fitting formula (Equation of (5) in BY19). It should be pointed out that the simulations of YB19 are calculated by considering the stars' gravity. 
As a result of the presence of outflow in natural systems, the black hole accretion rate is significantly reduced. Since the Bondi solution does not involve any rotation or wind and is quite simple, all of our numerical solutions and simulations of BY19 are lower than the Bondi accretion rate.

\section{Discussion and Concluding Remarks}\label{3}

In this study, we aim to investigate axisymmetric, steady, and viscous accretion flows with small, non-zero specific angular momentum. We have obtained several functional numerical solutions that completely describe the structure of slowly rotating accretion flows. We have neglected radiative cooling in our study. Also, for simplifying, we assume the flow dynamics is a one-temperature plasma, i.e., ion and electron have the same temperature, but in reality, there is two-temperature accreting plasma. The key element of the model is that accreting gas has a low angular momentum at the outer boundary. The low angular momentum accretion onto a black hole is common in nature. Observations of elliptical galaxies have provided one surprise: in the densest region of massive elliptical galaxies, the hot gas rotates slowly \cite{1977ApJ...218L..43I}. To find transonic global solutions, we adopted a two-point boundary value problem. In this study, to maintain hydrostatic equilibrium along with vertical direction, we assume the initial state $H\neq r$ rather than the simple approximation $H = r$. We have also investigated the effects of the star's gravity and the outer boundary conditions in the accretion process and found that it plays an important role. 

\begin{itemize}

\item Our results demonstrate both slight rotations and maintaining hydrostatic equilibrium reduce the mass accretion rate onto the black hole. Furthermore, taking into account a vertical hydrostatic equilibrium
the equation will alter the position of the sonic point.

\item The mass accretion rate is slightly enhanced when the potential of the galaxy is considered. In our study, the mass accretion rate increases when we make the viscosity larger. The viscosity accelerates the accretion process.

\item As the temperature of the outer boundary increases, the mass accretion rate decreases slightly. Therefore, the external boundary conditions may affect the mass accretion rate. 

\item Moreover, the transition from Bondi-like to Disk-like is discontinuous and well shown for inviscid and adiabatic flows by \cite{1981ApJ...246..314A}, \cite{1989ApJ...336..304A}. In this work, similar to previous results, we see that this transition occurs in viscous flows (\cite{1999ApJ...521L..55Y}, \cite{2001AIPC..556...93Y}, NF11).

\end{itemize}
In this study, the effects of outflow, energy convection, and MHD have been neglected just for simplicity.
We now have a better view of the behavior of slightly rotating accretion flows. Furthermore, we know
the outflow produced from hot accretion flow has been confirmed by observations and simulations, and its presence is undeniable. Outflow can significantly influence the dynamics of accretion flow around the black hole. The effect of the outflow on the global solution of the hot accretion flow is beyond the scope of this paper, and we postpone it to our future study.

\section*{ACKNOWLEDGMENTS}
We would like to thank Santabrata Das for his useful discussions. The authors thank the anonymous referees for the careful reading of the manuscript and their insightful and constructive comments. We hereby acknowledge the Sci-HPC center of the Ferdowsi University of Mashhad where some of this research was performed. A.M. is supported by the National Natural Science Foundation of China (Grant No. 12150410308), the foreign expert's project (Grant No. QN2022170006L), and also the China Postdoctoral Science Foundation (grant No. 2020M673371). Also, we have made extensive use of the NASA Astrophysical Data System Abstract Service. This work was supported by the Ferdowsi University of Mashhad under grant no. 57030 (1400/11/02).

\section*{DATA AVAILABILITY STATEMENT}

No new data were generated or analyzed in support of this research






\bsp	
\label{lastpage}
\end{document}